# A Bayesian dawn in linguistics: Trends, benefits and good practices

Natalia Levshina

Centre for Language Studies, Radboud University

**Abstract**

In recent years, Bayesian statistics has gained traction across a wide range of scientific disciplines. This paper explores the growing application of Bayesian methods within the field of linguistics and considers their future potential. A survey of articles from different linguistics journals indicates that Bayesian regression has transitioned from fringe to fairly mainstream over the past five years. This paper discusses the main drivers of this shift, including the increased availability of user-friendly software and the replicability crisis in adjacent disciplines, which exposed the shortcomings of the traditional statistical paradigm. It outlines the fundamental conceptual distinctions between frequentist and Bayesian approaches, emphasizing how Bayesian methods can help address the problems. Additionally, the paper highlights the methodological benefits of Bayesian regression for a diverse array of research questions and data types. It also identifies key theoretical and practical challenges associated with Bayesian analysis and offers a set of good practices and recommendations for researchers considering the adoption of Bayesian methods.

## 1. Bayesian statistics and its role in the history of science

In some respects, the history of Bayesian statistics parallels that of Artificial Intelligence (AI). Like AI 'springs' and 'winters', Bayesian methods have experienced cycles of enthusiasm and decline. Today, both fields are enjoying renewed interest and popularity. In both cases, this revival has been largely driven by advances in computational power and algorithmic innovations, such as convolutional networks and transformer architectures for AI, and Monte Carlo Markov Chain methods like Gibbs and Hamiltonian sampling for Bayesian statistics.

The origins of Bayesian statistics trace back to Reverend Thomas Bayes, who in the 18th century formulated a special case of what we know now as Bayes rule. However, it was Pierre Simon Laplace who formalized the rule in its modern form and applied it to a range of scientific problems. In this way, Bayesian statistics is an example of Stigler's law of eponymy, which says that no scientific discovery is named after its original discoverer (Stigler 1980). Bayesian statistics largely dominated the statistical practice throughout the 19th century but were supplanted by frequentist approaches in the 20th century (Efron 2005).

For much of its history, Bayesian statistics was referred to as "inverse probability", reflecting its focus on reasoning backward from observed data to unknown parameters (Dale 1999). Ironically, one of the earliest uses (if not the first) of the adjective "Bayesian" came from Sir Ronald Fisher, who employed it in a dismissive tone (Fienberg 2006). Despite that, the term was embraced by the proponents of the approach, as this happens with reclaimed slurs. Today, "Bayesian" is used neutrally by both advocates and critics of the methodology.

As for "frequentist", or traditional, non-Bayesian statistics, the label stems from its interpretation of probability as the long-run frequency of events in repeated experiments. Frequentist methods include p-values developed by Ronald Fisher, the hypothesis testing framework introduced by Jerzy Neyman and Egon Pearson, and Neyman's concept of confidence intervals. The central tool of inference in this framework is known as Null Hypothesis Significance Testing (NHST). Interestingly, much like the term "Bayesian", the label "frequentist" only began to gain traction in the 1950s, primarily among Bayesians seeking a term to describe their methodological counterparts (Fienberg 2006). In both cases, the labels were originally coined to describe the "other side", often in a non-complimentary way.

Bayesian statistics declined in the first half of the 20th century due to a combination of conceptual, practical and political factors. Conceptually, it was closely tied to the interpretation of probability as subjective belief, often referred to as "subjective" or "personalistic" probability. This was against the intellectual climate of that period, which found its reflection in logical positivism and its emphasis on objectivity and empirical verification. Anything associated with subjectivity or "beliefs" was viewed as metaphysical and unscientific. Practically, Bayesian methods were hindered by computational limitations. As statistical problems grew more complex, calculating posterior distributions became intractable. Political dynamics also played a role, as it is common in science. Frequentist statistics dominated the US statistics departments - particularly through the influence of Jerzy

Neyman and his colleagues at Berkeley, as well as through Ronald Fisher's enduring impact on U.S. land-grant universities with focus on agriculture (Fienberg 2006).

In the middle of the 20$^{th}$ century, Bayesian statistics experienced a notable revival, driven by the mathematical and philosophical contributions of such statisticians as Irving John Good, Leonard J. Savage and Bruno de Finetti. A critical development was the creation of Monte Carlo methods, which are crucial for Bayesian inference, by John von Neumann, Stanislaw Ulam, and Nicholas Metropolis. However, these methods only gained widespread application in different scientific domains later, with the advent of more powerful computers and efficient algorithms.

A milestone came in 1989 with the development of *Bayesian inference Using Gibbs sampling* (BUGS), a Bayesian software that made it feasible to perform Monte Carlo Markov Chains computations. BUGS became a widely adopted tool in the sciences. Its Windows version WinBUGS (released in 1997) became especially popular, accounting for nearly half of all Bayesian empirical publications up to 2012 (van de Schoot et al. 2017). By enabling Bayesian computations on standard desktop machines, it significantly broadened access to these methods. The release of C++ software like JAGS (Just Another Gibbs Sampler, first released in 2007), and Stan (an open-source package for obtaining Bayesian inference using the No-U-Turn sampler, a variant of Hamiltonian Monte Carlo, first released in 2013), made Bayesian methods even more efficient and accessible. These tools support multiple platforms and offer integration with R through packages like rjags and Rstan. However, their use required some programming skills, which posed a barrier for many applied researchers.

The release of the *brms* package in 2017 (Bürkrner 2017), a wrapper for Stan with syntax familiar to the popular *lme4* package for mixed-effects modelling, made Bayesian regression more accessible to users with little or no programming background. People preferring a drag-and-drop interface like in SPSS, can do Bayesian analyses in JASP (JASP Team 2025).

Beyond the advances in software, the intellectual climate has also been favourable for Bayesian methods, through the development of Bayesian theories of perception, learning, reasoning and language processing in cognitive science (cf. Kruschke 2010). These theories and models, which apply Bayes rule to formally describe how humans use prior beliefs or expectations when processing new information and how these beliefs are constantly updated in light of new evidence (e.g., Chater et al. 2010), familiarized researchers with Bayesian thinking and helped legitimize Bayesian statistical methods.

Another significant factor was the replicability crisis in adjacent disciplines, particularly within social sciences. The crisis exposed the inability to reproduce many published findings, highlighting deeply rooted issues in traditional statistical practices, especially the reliance on NHST and *p*-values, which has been linked to a range of problems, from misinterpretation of results (Goodman 2008) to questionable statistical practices, such as p-hacking and HARKing (Hypothesizing After the Results are Known), as well as broader issues, such as publication bias (Stefan & Schönbrodt 2023). Within linguistics, the implications of the replicability crisis have received attention in numerous papers and special issues (e.g., Sönnig & Werner 2021; Schmalz et al. 2025).

The awareness of the poor practices prompted efforts in all disciplines, including linguistics. Those include the promotion of open data and code, preregistration and overall greater methodological transparency. In a notable response, the journal *Basic and Applied Social Psychology* even imposed a ban on p-values and confidence intervals in all submissions, arguing that "the $p < .05$ bar is too easy to pass and sometimes serves as an excuse for lower quality research" (Trafimow & Marks 2015: 2). The journal instead encouraged its authors to focus more on descriptive statistics and increase sample sizes.

Amid these broader developments, Bayesian statistics emerged as a viable alternative to the traditional approaches. According to a survey by Lenhard (2022), the proportion of papers related to Bayesian statistics in five major statistical journals rose from under 5% in 1990 to approximately 20% by 2015.

In social sciences, the adoption of Bayesian statistics has been more modest. Lynch & Bartlett (2019) report that economics has been the most receptive, with the share of Bayesian papers increasing from 0.03% in the 1990s to about 1.5% in 2008, after which it stabilized. Other disciplines have shown slower uptake: by 2017, less than 1% of all papers in psychology were Bayesian, with even more modest numbers observed in sociology and political science.

Despite these relatively small figures, the number of Bayesian papers continues to grow. A literature review by van de Schoot et al. (2017) identified over 1,500 Bayesian articles in Psychology published between 1990 and 2015, with a clear upward trend across that period. Notably, Bayesian regression methods were used nearly half of those studies.

In 1975, British statistician Dennis Lindley predicted that the 21st century would be Bayesian (Lindley 1975). While this vision has not been fully realized, Bayesian methods have undeniably gained significant traction. These developments have also begun to influence linguistics and psycholinguistics. Roughly a decade ago, the first Bayesian regression models started appearing that used Stan or JAGS. One early example is Vasishth et al. (2013), who used JAGS to conduct a Bayesian meta-analysis of studies on the processing of Chinese relative clauses. Shortly thereafter, Bayesian multilevel linear models were applied to data from individual experimental studies. In 2014, Vasishth's lab reported two such applications for modelling reading times: Husain et al. (2014) on Hindi, and Hofmeister and Vasishth (2014) on English. Other early applications include Gudmestad, House, and Geeslin (2013), who used Bayesian multinomial probit regression to model subject expressions in L1 and L2 Spanish, and Skrivner (2015), who fitted a generalized mixed-effects logistic model in JAGS to investigate VO/OV order in Latin and Old French.

As Nicenboim et al. (2018: 46) optimistically observed, "in the last few years, there has been a strong move towards Bayesian modeling" in linguistics and psycholinguistics. But how popular are actually Bayesian models in linguistics? This is the central question addressed in Section 2, which examines the Bayesian shift based on data from linguistics journals. The rest of the paper is structured as follows. Section 3 introduces the basic concepts of Bayesian concepts and contrasts them with their frequentist counterparts, highlighting the ways in which Bayesian methods can help address the replicability crisis. Section 4 explores the practical benefits of Bayesian regression and the use of Bayes Factor for a wide range of research questions and data types in linguistics. Finally, Section 5 offers guidance for researchers considering whether to adopt Bayesian techniques, along with a set of best practices for their effective implementation.

## 2.    Bayesian regression in linguistics: state of the art

Kortmann (2021) identifies a "quantitative turn" in linguistics, which occurred between the end of the 20th and the beginning of the 21st century. With some variation in timing, this shift has impacted many subfields. For example, 2008 was a pivotal year in the history of Cognitive Linguistics. According to Janda (2013), prior to that year, most articles in the field were qualitative. After 2008, however, a majority began to incorporate some sort of quantitative analysis based on authentic language data. This raises a question: has there also been a similar "Bayesian turn"?

Although one can occasionally see other Bayesian methods in linguistic papers (e.g., Bayesian Kendall's *tau* in Milin et al. 2023), Bayesian regression is by far the most popular. This is why this section zooms in on the use of Bayesian regression compared to traditional regression and other methods. I manually examined articles published in ten linguistic journals in 2019 and 2024, focusing on empirical studies and omitting book reviews, short commentaries and position papers without empirical data. I reviewed a total of 243 papers from 2019 and 253 from 2024 and classified them into three groups: a) papers reporting a Bayesian regression model, b) articles using a traditional regression model (usually estimated with the help of maximum likelihood methods), excluding ANOVA, and c) all other studies. Although most articles do not explicitly state that they use traditional regression, this information could be inferred from references to R packages, such as *lme4* or *lmerTest*, functions, such as *glm()*, or the reporting of *p*-values. In a small number of cases, where this information was not available, I assumed that traditional statistics was used.

Figure 1 shows the proportions of Bayesian and maximum likelihood regression use in 2019 and 2024. The overall proportion of papers reporting any kind of regression analysis has increased from 34.1% in 2019 to 46.6% in 2024. Bayesian regression has also become more common: only 1 paper (less than 1%) used it in 2019, compared to 19 papers (7.5%) in 2024. As a result, the proportion of Bayesian approaches among all regression papers has increased substantially, from 1.2% in 2019 to 16.1% in 2024. This is a noticeable change in statistical practices.

Table 1 presents the distribution of Bayesian and any kind of regression models across the ten journals in 2024. As one can see, papers with Bayesian regression appear in most of the journals, indicating that its adoption is widespread rather than concentrated within specific journals or linguistic subdisciplines. This suggests that Bayesian regression is gradually gaining acceptance across the broader field of linguistics.

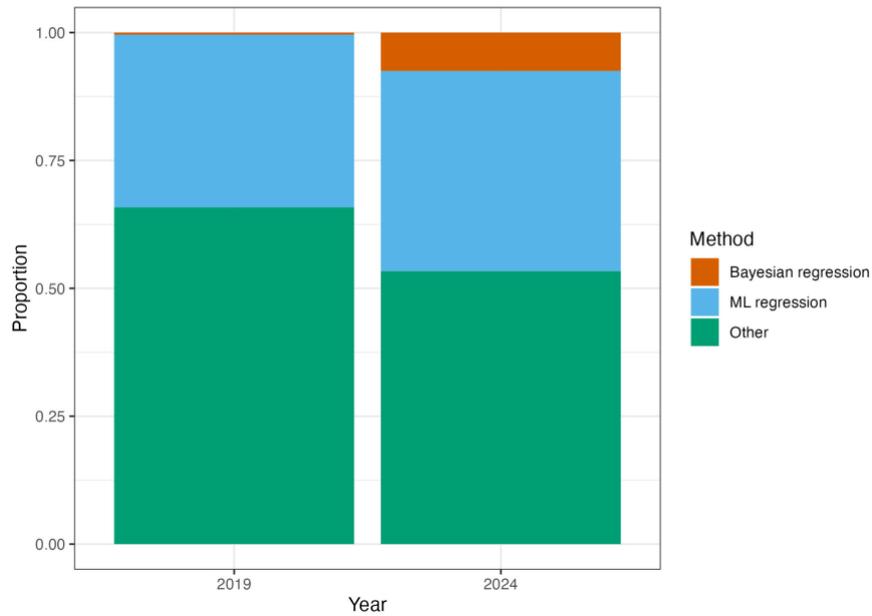

Figure 1. Proportions of papers reporting Bayesian and maximum likelihood (ML) regression models in ten linguistic journals in 2019 and 2024.

Table 1. Papers reporting Bayesian and other regression in 2024 in ten linguistic journals.

| Journal | N papers checked | N papers reporting any kind of regression | N papers reporting Bayesian regression |
|---|---|---|---|
| Cognitive Linguistics | 19 | 11 | 4 |
| English Language and Linguistics | 24 | 6 | 1 |
| International Journal of Corpus Linguistics | 20 | 3 | 0 |
| Journal of Phonetics | 22 | 20 | 3 |
| Journal of Pragmatics | 25 | 4 | 0 |
| Language | 22 | 11 | 2 |
| Language and Cognition | 50 | 14 | 2 |
| Language Variation and Change | 15 | 13 | 3 |
| Linguistic Typology | 13 | 3 | 2 |
| Linguistics | 43 | 15 | 2 |
| *Total* | *253 (100%)* | *118 (46.6%)* | *19 (7.5%)* |

To conclude, we do not see a full Bayesian turn in linguistics yet, but we do observe a Bayesian "dawn". Why is it happening now? The adoption of any new information technology, including statistical methods, depends on several key factors outlined in technology acceptance theories (Davis 1989; Venkatesh et al. 2003):

- Performance expectancy: the perceived benefits, such as usefulness, productivity, or impact on one's career.
- Effort expectancy: the perceived ease of learning and using the method;
- Social influence: the perception that influential people, institutions or one's own organization support the use of the system.

When researchers perceive Bayesian methods as useful, accessible, and observe growing adoption within their academic community, they are more likely to form an intention to adopt these methods. This intention often translates into actual usage. Are all these conditions satisfied in linguistics?

Let us begin with performance expectancy. Many language scientists who use Bayesian regression in their papers argue that it offers advantages over traditional methods. One of the most frequently cited benefits is the ability to evaluate the posterior probability of certain values (e.g., positive or negative) of a regression parameter. This allows researchers to make direct statements about their research hypothesis, rather than merely trying to reject a null hypothesis (e.g., Husain et al. 2014). The conceptual foundations and advantages of Bayesian methods in comparison with frequentist approaches are discussed in Section 3.

In addition, the probabilistic nature of Bayesian inference and rich information provided by posterior distributions make it particularly well-suited for variational and usage-based linguistics. For example, Bayesian regression is a perfect match for probabilistic approaches to language variation because posterior probabilities can be easily compared across languages or language varieties (e.g., Levshina 2018). It is also more suitable for many specific research questions and types of data - see an overview in Section 4.

Sometimes the choice of Bayesian methods has been motivated by more pragmatic considerations, such as the lack of readily available software for equivalent frequentist approaches. For instance, hierarchical models with a categorical response variable with more than two outcomes were fitted using Bayesian regression studies by Gudmestad et al. (2013) and Levshina (2016), primarily because no accessible frequentist alternatives could be found at the time.

With regard to effort expectancy, many linguists may perceive learning yet another statistical framework as too demanding. However, if one already has experience with traditional regression methods, transitioning to Bayesian regression is relatively straightforward. This likely explains the observed correlation between the use of regression techniques and the proportion of Bayesian regression in linguistic papers. As already mentioned, s key turning point has been the development of the *brms* package (Bürkner 2017), which mirrors the syntax of *lme4* and enables researchers to use Stan without acquiring additional programming skills.

The learning of Bayesian methods is supported by numerous accessible resources, including textbooks (e.g., Kruschke 2015; McElreath 2016; Kaplan 2024; Nicenboim et al. 2026), conceptual primers (Kruschke 2010; Van Zyl 2018; Pek & Van Zandt 2020; van de Schoot et al. 2021; Levshina 2022), and practical tutorials (Wagenmakers et al. 2010; Nicenboim and Vasishth 2016; Sorensen et al. 2016; Vasishth et al. 2018; Winter & Bürkner 2021). In addition, the *brms* package offers excellent vignettes covering a wide range of models. Introductory courses and workshops in Bayesian statistics have also been offered in summer and winter schools for early-career researchers.

Another important aspect is the ease of usage. Although Bayesian algorithms have improved significantly in recent years, Bayesian methods often remain more time-consuming and computationally intensive than their frequentist counterparts, especially when applied to complex models or large datasets. This is due to the reliance on iterative sampling from the posterior distribution. Additionally, setting up the necessary infrastructure, such as installing Stan and configuring the C++ toolchain, can represent a technical challenge for some users.

One commonly perceived challenge in Bayesian analysis is the choice of priors. However, this concern is often overstated. In most cases, researchers can select reasonable priors without much difficulty, based on common sense and previous knowledge. For example, an effect of an input variable in a logistic regression model is rarely more than 5 on the logit scale, which corresponds to the change in probability from 0.5 to 0.99 or from 0.01 to 0.5; and it is extremely unlikely to be more than 10, which moves the probability from 0.01 to 0.99 (Gelman et al. 2008). In practice, priors are often chosen pragmatically, as a regularization tool that enhances model performance model (Lenhard 2022). In any event, one should conduct sensitivity analyses to assess the impact of different priors on the posteriors.

On the other hand, some researchers may underestimate the complexity of applying Bayesian methods. A common misconception is that Bayesian regression requires little to no diagnostics and is free from convergence issues. This may lead to an overly optimistic estimation of the effort involved. In reality, both Bayesian and frequentist approaches require thorough model diagnostics and criticism. In the Bayesian context, one should verify that the Markov chains have reached their stationary distribution, using diagnostics such as the R-hat statistic and effective sample size. Posterior predictive checks are also crucial for assessing if the simulations from the fitted model adequately represent the observed data (Gelman et al. 1996). Section 5 discusses these and other good practices of conducting a Bayesian analysis.

Finally, the adoption of Bayesian methods in linguistics also depends on peer influence and community acceptance. The analysis of ten journals suggests that journal editors and reviewers generally do not hinder the spread of Bayesian approaches.

## 3. Main differences of Bayesian statistics from traditional statistics

3.1. Probability as frequency vs. probability as belief

In frequentist statistics, an event is considered probable if it is expected to occur often in repeated trials. For example, if a coin is fair, and we toss it many times, we expect half of the tosses to be heads and half to be tails. The more times we repeat the experiment, the closer the observed proportions of heads and tails will be to 50% each. In this view, probability can be defined as a long-run frequency. This is why traditional statistics is called "frequentist".

In Bayesian statistics, probability is usually interpreted as a measure of belief – a broad term that covers expectations, opinions, assumptions, convictions and so on. For any given value of a parameter or a range of values, we can assign a probability.

In a sense, Bayesian probabilities exist in our heads. But this can also be said about frequentist probabilities. After all, the notion of an infinite number of identical repetitions requires imagination, too.

3.2. Parameters as point estimates vs. parameters as probability distributions

Frequentist statistics assumes that that a population parameter is an unknown but fixed constant. If an experiment is repeated an infinite number of times, the results would converge to the true value of that parameter. For instance, with a a fair coin tossed an enormous number of times, exactly 50% of tosses will be heads (and 50% tails) – a point estimate. Bayesian statistics, in contrast, represents unknown parameters as random variables described by probability distributions. Some values are considered more plausible than others based on the data and prior beliefs.

In practical terms, when you fit a traditional regression model, the intercept or slope coefficient is reported as one number. In contrast, if a Bayesian regression is fitted, you obtain a whole probability distribution. For example, Figure 2 shows a hypothetical distribution for a regression coefficient representing the effect of time (in years) on linguistic diversity, measured as the number of existing first (L1) languages. According Bromham et al. (2022), at least one language is lost per month without intervention – roughly 12 languages per year. This value corresponds to the peak (the mode) of the distribution in the figure, as indicated by the highest probability density on the y-axis. This is also the mean and the median of this distribution. However, the bell-shaped curve also allows for other plausible values, such as -15 or -10 languages per year. More extreme values, like -25 languages per year, or no loss at (0 languages per year), have much lower probability.

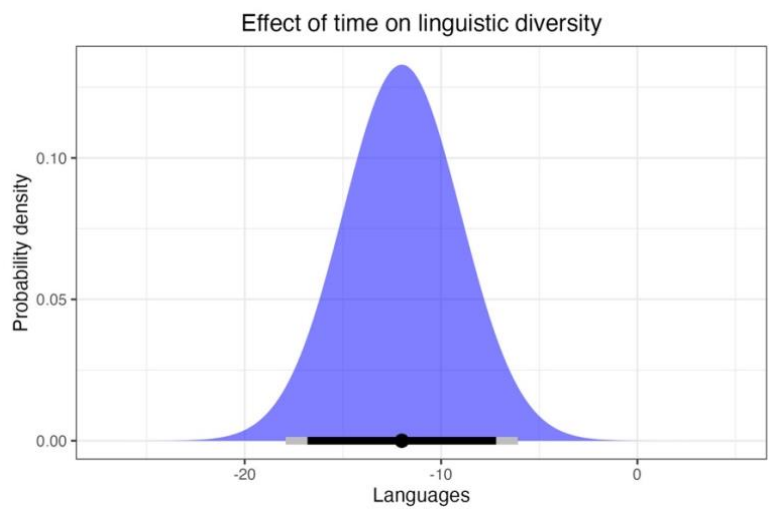

Figure 2. Imaginary probability distribution of an effect of time (in years) on linguistic diversity. The dot represents the mode, which in this case coincides the the mean and the median ($x = -12$). The black line shows the 89% credible interval, whereas the gray line behind it corresponds to the 95% credible interval.

We typically summarize a posterior distribution by reporting its mean along with a credibility interval (explained below). In Figure 2, the gray line shows the 95% credibility interval, while the black one is the 89% credibility interval, following McElreath's (2016) suggestion to use less trivial cut-off points. Other summary statistics are also possible, such as the median or the mode (also called the "maximum a posteriori" estimate). In this example, all these measures coincide with the mean. But what is a posterior distribution and how to we obtain it? These questions are addressed below.

3.3. Priors and posteriors

Bayesian statistics is essentially the application of Bayes' rule – a fundamental principle that is neither inherently Bayesian nor frequentist – to the process of hypothesis testing. Bayes rule concerns the computation of conditional probabilities, denoted by $P(a|b)$, which represents the probability of event a given that event b has occurred. For example, the probability of having coronavirus given a positive test result can be expressed as follows:

(1) $\quad P(\text{virus} \mid \text{positive}) = P(\text{positive}|\text{virus}) * P(\text{virus}) / P(\text{positive})$,

where $P(\text{positive}|\text{virus})$ is the probability of testing positively given that the person has the virus – also known as the true positive rate. $P(\text{virus})$ is the probability of having the virus in the population, often referred to as the infection rate. Finally, $P(\text{positive})$ is calculated as the sum of the probability of a true positive, and the probability of a false positive.

Bayesian statistics begins when we apply Bayes' rule to epistemic beliefs. Suppose we want to determine the probability distribution of a certain parameter $\theta$ given some experimental or corpus data, $P(\theta|\text{Data})$. This is called the **posterior probability distribution**, or simply the posterior. It represents the degree of belief in each possible value of the parameter after taking the data into account.

The distribution can be obtained from three components:

1. **Prior probability** $P(\theta)$, or simply the prior. This is the probability distribution of parameter θ before observing the data.
2. **Likelihood** $P(\text{Data}|\theta)$: the probability of observing the data given a specific value of θ.

3.  **Marginal likelihood** *P* (Data), or evidence: the probability of the data. It is usually a normalizing constant that helps make sure that the posterior distribution integrates to one.

(2)     *P* (θ| data) = *P* (data|θ) * *P* (θ) / *P* (data), or simply

(3)     Posterior distribution = Likelihood x Prior distribution/Evidence

Because Evidence is merely a normalizing constant, we can express this as:

(4)     Posterior distribution ∝ Likelihood  x  Prior distribution

where the symbol ∝ means 'proportional to'.

Consider an illustration. Let us imagine three researchers, each independently studying the effect of word surprisal on an eye-movement metric in a reading experiment. The metric, regression-path duration, is the sum of all fixations beginning with the initial fixation on a word and ending when the reader's gaze is directed away from the region to its right. It is measured in milliseconds. Surprisal, which represents the relative unexpectedness of a word in context, is associated with greater processing complexity and longer reading times (Hale 2001; Demberg & Keller 2008; Levy 2008; Lowder et al. 2018, *inter alia*). It is measured in bits and computed as negative log-transformed probability of a word given its preceding context.

One researcher – let's call her Assertia – has strong prior beliefs about the effect of surprisal. Based on earlier work, she expects a positive effect of around 10 ms per bit, with little room for variation. Such priors are called informative. Now suppose that the data strongly suggest an effect of 20 ms per bit, as indicated by the Likelihood curve in Figure 3. The resulting posterior distribution would fall between the prior and the likelihood, reflecting both Assertia's expectations and the new evidence.

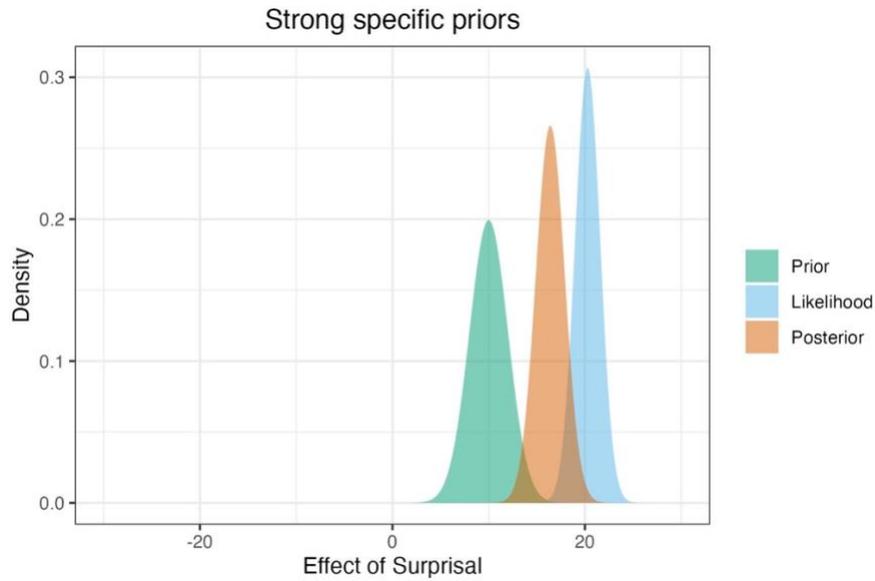

Figure 3. Effect of Assertia's strong priors on the posteriors of the effect of suprisal on regression-path duration.

The next researcher, who we will call Brelax, has no expectations about the effect of surprisal. He adopts uniform (or flat, diffuse) priors, meaning that no parameter values are given preference over others. Such priors are often referred to as non-informative priors, though some statisticians consider this a misnomer: even a completely flat prior still conveys information about the degree of uncertainty (van de Schoot et al. 2021). Flat priors "let the data speak for itself", which may be desirable in certain settings, such as clinical studies (Kaplan 2024: 18). The data analyzed by Brelax is identical to Assertia's. When diffuse priors are used, the posterior distribution coincides with the likelihood, as shown in Figure 4.

A common misconception is that Bayesian statistics is entirely different from frequentist statistics. In reality, the two share very many similarities - most notably, the use of the likelihood function. If we adopt flat priors, the posterior becomes proportional to the likelihood, differing only by a normalizing constant:

(5)     Posterior $\propto$ Likelihood

In such cases, using the posterior mean yields results that are virtually identical to those from a frequentist analysis. The same holds when we have large datasets, where the influence of the prior becomes negligible. As the sample size increases to infinity, the result becomes the same as what we would have obtained with the help of the frequentist method. This fact is known as the Bayesian central limit theorem.

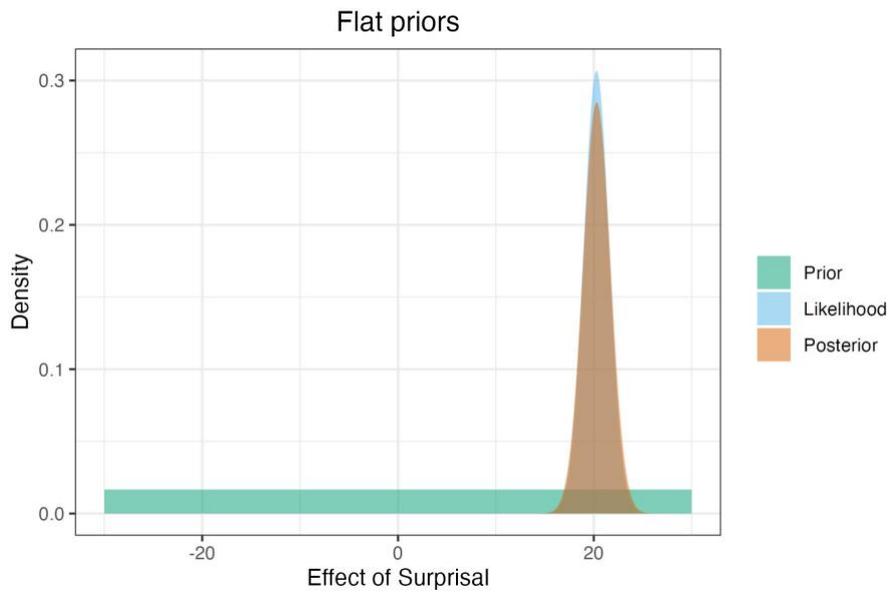

Figure 4. Effect of Brelax' flat priors on the posteriors of the effect of suprisal on regression-path duration.

Finally, consider the third researcher, Calibra (from "calibrate"). Like Assertia, she specific priors centred around 10 ms per bit, but her priors are less restrictive: the distribution is more spread out, allowing for greater variation. Calibra analyzes the same data as Assertia and Brelax. As shown in Figure 5, these weaker priors exert only a minimal effect on the posterior distribution, which remains very close to the likelihood. Such priors are often called weakly informative. They are commonly used for regularization, helping to stabilise model estimates without materially influencing the resulting inferences (Kaplan 2024: 20-21).

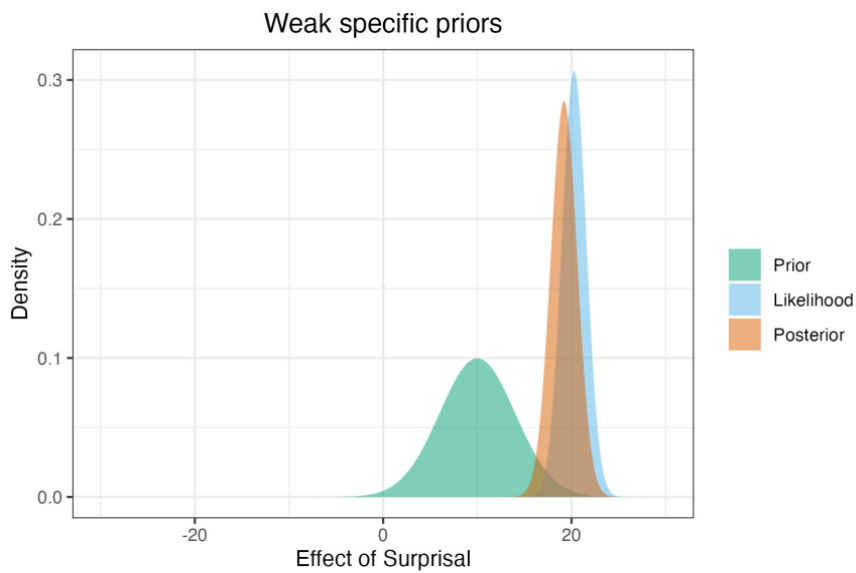

Figure 5. Effect of Calibra's weakly informative priors on the posteriors of the effect of suprisal on regression-path duration.

## 3.4. Estimation of parameters

In classical regression, parameters are typically estimated using ordinary least squares or maximum likelihood methods, which rely on assumptions about the underlying data distributions. In most Bayesian applications, posterior distributions are too difficult to derive analytically, which made Bayesian statistics impractical for much of its history. As noted in Section 1, Bayesian regression became feasible with the advent of Monte Carlo Markov Chain (MCMC) methods, which combine Monte Carlo sampling with Markov chains.

Monte Carlo methods rely on the law of large numbers. If you spend enough time at a roulette table, each number from 0 to 36 will appear about 1/37 of the time. Although it may sound trivial at first, this fact makes it possible to approximate any underlying distribution by drawing many samples from it, turning a complex integration problem to a simple iteration problem (Lenhard 2022). However, sampling blindly from a vast range of values can be too slow even for powerful computers. This is where the second MCMC component, Markov chains, can help. A Markov chain is a random walk in which the next step depends only on the current position and a probability of moving to the next position. By restricting each move to a smaller set of possibilities, the chain efficiently explores the space: some values are visited more often, others less. Crucially, Markov chains converge to a stationary distribution (or reach an equilibrium), regardless of the starting point. [1]

How does this work in Bayesian regression? Suppose we need to estimate a coefficient. An MCMC algorithm begins with a random starting value for that coefficient and then takes a guided random walk through the parameter space, informed by both the priors and the data. Over time, the walk spends more time in regions where values are more probable and avoids less likely regions. The values "visited" during this walk are called posterior samples. Once the Markov chain has converged to its stationary distribution, these samples approximate the posterior distribution. We can then compute the mean, the median and other useful statistics directly from this sequence of sampled values.

## 3.5. Null Hypothesis Significance Testing vs. Bayesian hypothesis testing

How does all this help us test research hypotheses? In frequentist statistics, the standard approach is null hypothesis significance testing (NHST). The null hypothesis typically states that there is no correlation, no difference, or no effect of a variable on the response. This is evaluated using the *p*-value – the probability of obtaining the observed test statistic and more extreme values, assuming that the null hypothesis is true. Conventionally, the null hypothesis is rejected if the p-value is smaller than 0.05, and the alternative is then accepted.

---

[1] Note that there exist Bayesian methods without MCMC, such as sequential Monte Carlo and approximate Bayesian computation (see e.g., van de Schoot et al. 2021).

This procedure is routine in most disciplines, but is plagued by problems and misconceptions. In particular, *p*-values are notoriously difficult to interpret and often misunderstood (Goodman 2008). For example, a study of articles in top psychological journals found that 72% of papers with "null results" ($p > 0.05$) incorrectly inteprated them as evidence in favour of the null hypothesis (Aczel et al. 2018). Authors wrote statements, such as "The results establish that the intervention has no effect on the dependent variable", or "There was no difference between the participants in in the intervention group and the control group". This is incorrect: a large *p*-value does not mean there is a high probability that the null hypothesis is true (nor that the alternative hypothesis is false).

These conceptual difficulties are not surprising, if one realizes that NHST is a hybrid of two distinct statistical approaches. The first is Ronald Fisher's use of *p*-values to indicate how well the data fit the null hypothesis without making decisions about the alternative. The second comes from Neyman–Pearson decision theory, which focuses on balancing the the risks and costs of Type I errors (false positives) and Type II errors (false negatives) to choose between the null and alternative hypotheses. This framework is about decision-making under risk, not about determining which hypothesis is true or how well the data support either one (Pek & Van Zandt 2020).

Bayesian statistics does not rely on *p*-values, avoiding the binary "significant/not significant" or "accept/reject" mindset. But what should we use instead? Bayesian analysis offers several alternatives: Bayes factors, Regions of Practical Equivalence (ROPE), Maximum A Posteriori-based *p*-values, *e*-value in the Full Bayesian significance test (FBST), and the Probability of Direction. However, there is no universal agreement on which measure to prefer, and each has its strengths and limitations, which may be a challenge for beginners (cf. Makowski et al. 2019; Kelter 2020).

For illustration, consider Probability of Direction (PD), which is simply the proportion of the posterior distribution sharing the sign of the median. In Figure 6, almost the entire distribution on the left is negative, giving a PD close to 100%. This suggests strong support for the negative effect. On the right, 90.5% of the distribution is positive, indicating some evidence for the effect but not overwhelmingly so. PD is appealing because it is intuitive and easy to compute. However, it has a limitation: even when no effect exists, PD can hover around 0.6–0.7 rather than near 0.5, meaning it does not always clearly signal the null case (Kelter 2020). If it is necessary to quantify the support for the null hypothesis, one should use other measures, such as Bayes Factor.

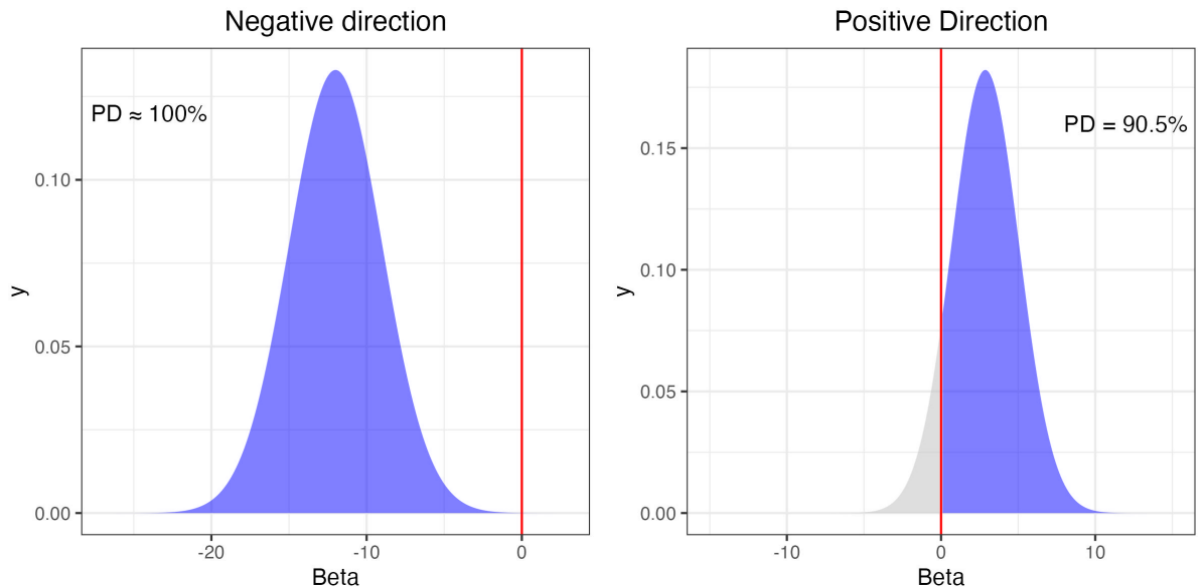

Figure 6. Probability of Direction (PD) as a measure that helps test if an effect is present.

3.6. Confidence vs. credible intervals

To convey uncertainty about estimated paramters, frequentist statistics uses 95% **confidence** intervals, whose meaning is at least as tricky as that of *p*-values. Interpreted correctly, if we were to compute such an interval repeatedly from many samples, the true (fixed but unknown) mean would fall within its bounds 95% of the time. In other words, the interval reflects the long-run reliability of the method, and not the probability that the parameter lies within a specific range for a given dataset. Unsurprisingly, this subtle distinction is often misunderstood.

A Bayesian **credible** interval is far easier to interpret. A 95% credible interval simply means there is a 95% probability that the parameter of interest lies within that range. In other words, we can be 95% certain that the value falls in the interval – an interpretation many people mistakenly apply to confidence intervals in frequentist statistics.

There are two common types of credible intervals: the Equal-Tailed Interval (ETI) and the Highest Density Interval (HDI). For a 95% ETI, the bounds are set between the 2.5% and 97.5% quantiles of the posterior distribution. A 95% HDI, in contrast, is the narrowest interval containing 95% of the distribution's probability mass. When the distribution is symmetric, the two intervals coincide (as in Figure 2). But when the distribution is skewed, as in Figure 7, the 95% HDI may be shifted relative to the ETI. In such cases, the HDI is generally preferable because it captures the most probable values more effectively.

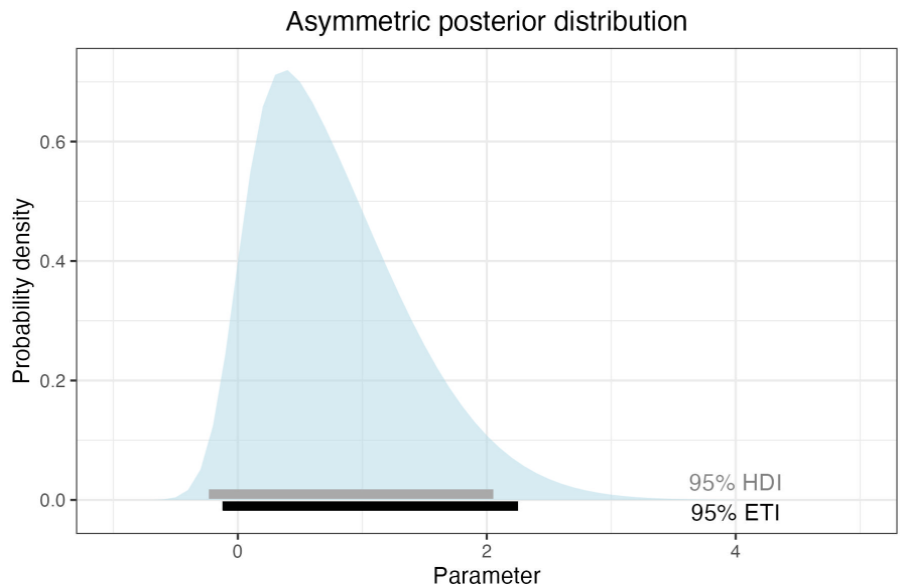

Figure 7. 95% Highest-density interval (gray) and 95% equal-tailed interval (black) of an imaginary asymmetric posterior distribution.

3.7. Conceptual advantages of the Bayesian approach

If Bayesian and frequentist analyses often produce similar results, one may ask, why bother with Bayesian methods? First of all, they have conceptual advantages, such as the possibility of making direct statements about parameters (e.g., there is a 99% probability that the parameter has a positive value), more intuitive interpretation of credible intervals, and integration of prior knowledge and common sense in statistical analysis.

Another key advantage is that Bayesian statistics promotes "a cumulative science that incrementally improves estimates of magnitudes and uncertainty" (Kruschke & Liddell 2018: 178; see also Schmid 1996). Cumulative science enables unbiased meta-analyses, and Bayesian inference can encourage the publication of evidence for the null hypothesis – something that the frequentist framework cannot do. This reduces the number of studies "in the file drawer" (Dienes 2014) and offers a path towards addressing the reproducibility crisis in many fields.

In addition to these conceptual reasons, for many linguists the main appeal lies in the flexibility of Bayesian methods. We are able to handle different types of data across a variety of situations, including challenging or problematic cases. These cases are discussed in next section.

## 4.   Special cases in which Bayesian statistics can be particularly useful

4.1. Complex random effects structure: Maximal models and phylogenetic regression

Bayesian regression offers an attractive approach for modelling complex random effects. For example, Husain et al. (2014), investigating expectation and locality effects on sentence processing in Hindi, aimed to fit maximal models in line with Barr et al.'s (2013) slogan, "Keep it maximal!". However, due to data sparseness, such complex models often fail to converge in *lme4*, and can yield unrealistic estimates of correlations between random effects (Bates et al. 2018). Using Stan, Husain et al. successfully fit a model for a 2-by-2 experimental design with random intercepts and slopes for both subjects and items, including an interaction term. Maximal Bayesian models have since been adopted in more recent psycholinguistic research (e.g., Copot & Bonami 2024; Scholman et al. 2024; Winkowski et al. 2024). Regularizing priors on correlations between random effects can also be employed to avoid degenerate and therefore useless estimates (Chung et al. 2013).

Another example is the incorporation of complex phylolinguistic information in cross-linguistic studies. Fine-grained data on genealogical relatedness between languages can be derived from phylogenetic trees, such as those available in Glottolog (Hammarström et al. 2020; https://glottolog.org ) or similar resources. This approach, known as hierarchical phylogenetic regression, originated in evolutionary biology (Hadfield & Nakagawa 2010) and is conveniently implemented in *brms* (Bürkner 2024). The method involves adding a covariance matrix of languages or other genealogical units to the random-effects structure. This matrix contains similarities between the selected phylolinguistic units, such that closely related units (e.g., German and Dutch) have values close to one, whereas unrelated languages (e.g., German and Japanese) have zero values. Incorporating such information addresses the problem of statistical dependencies between genealogically related languages, while retaining all languages in the sample and providing maximally detailed phylogenetic information (Guzmán Naranjo & Becker 2022).

## 4.2. Multinomial and ordinal mixed models

Stan (particularly via *brms*) makes it straightforward to fit mixed models with response variables containing more than two ordered or unordered categories. Such options are not available in the *lme4* package at the time of writing. Yet these data types are especially important in language variation studies and in experiments that employ Likert scales. Levshina (2016) used Bayesian multinomial regression in Stan due to the practical difficulties of applying traditional mixed-effects logistic regression to responses with more than two categories at the time. This can be done with the help of the 'categorical' family in *brms*.

For ordinal outcome variables, *brms* provides several modelling options (Bürkner & Vuorre 2019). Of particular relevance for psycholinguistic research are cumulative ordinal models, which are suited for Likert-scale data where ordered categories (presented as numbers or verbal labels) are assumed to reflect an underlying continuous psychological dimension, but other types can also be useful in particular applications. For example, in situations where the proportional odds assumption of cumulative models is violated, for instance, when some of the original categories are conflated due to data sparseness, one can use models with adjacent categories as a more flexible alternative (e.g., Hržica et al. 2024).

## 4.3. Data with many extreme values (zeros or ones)

Some datasets contain numeric data with many zeros and/or ones. This pattern often arises in experiments slider scales, where some participants prefer extreme values. Figure 8 illustrates such a distribution from de Hoop et al. (2023), who employed a 100-point slider scale to measure addressees' responses to emails written with either the informal T-pronoun or the formal V-pronoun in Dutch HR communication.

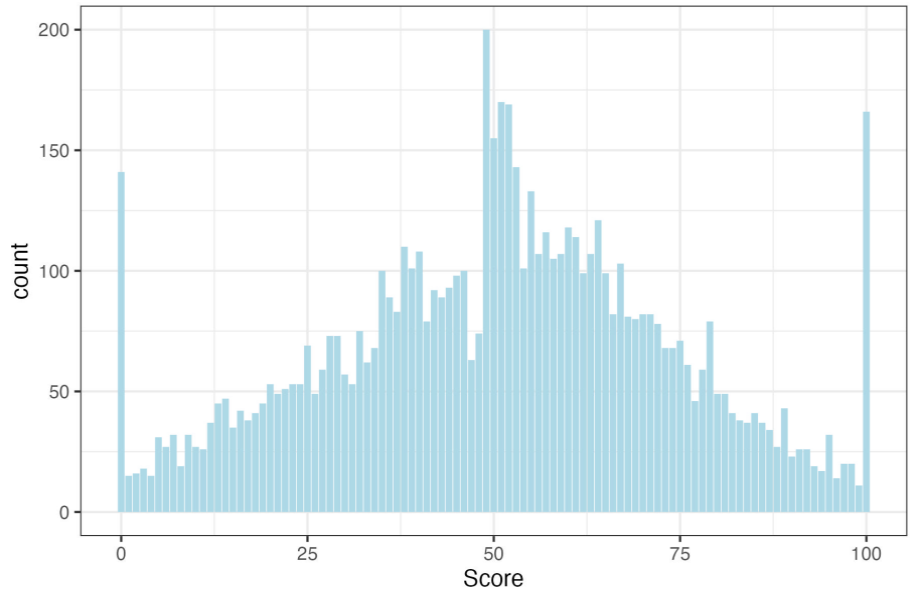

Figure 8. Data from an experiment with a 100-point slider scale (de Hoop et al. 2023).

If the data are transformed by dividing the scores by 100, the result is a typical case of zero-one inflation. Another complication is that the transformed scores are bounded between 0 and 1. A suitable solution is zero-one inflated beta (ZOIB) regression (Liu and Eugenio 2018), which can be conveniently implemented in Stan via *brms* using the "zero_one_inflated_beta" family. A ZOIB model combines several regression models:

- a standard beta regression for scores strictly between 0 and 1;
- a model of the *phi* parameter, which represents the precision, or inverse dispersion of the beta distribution;
- a logistic model predicting whether the response is 0 or 1 versus all other values;
- a logistic model predicting 1 versus 0, given that the outcome is at one of these extremes.

This approach offers flexibility in modeling variability in participants' behaviour. Specifically, how do they use the range of the scale? Do they prefer extreme values or values around the middle? As de Hoop et al. (2023) show, these preferences greatly across individuals. Figure 9 shows data from three participants in their experiment. Participant A uses the entire scale fairly evenly but avoids the extremes. Participant B prefers mid-range scores. Finally, Participant C selects only extreme values. This kind of individual variability can be captured using ZOIB regression.

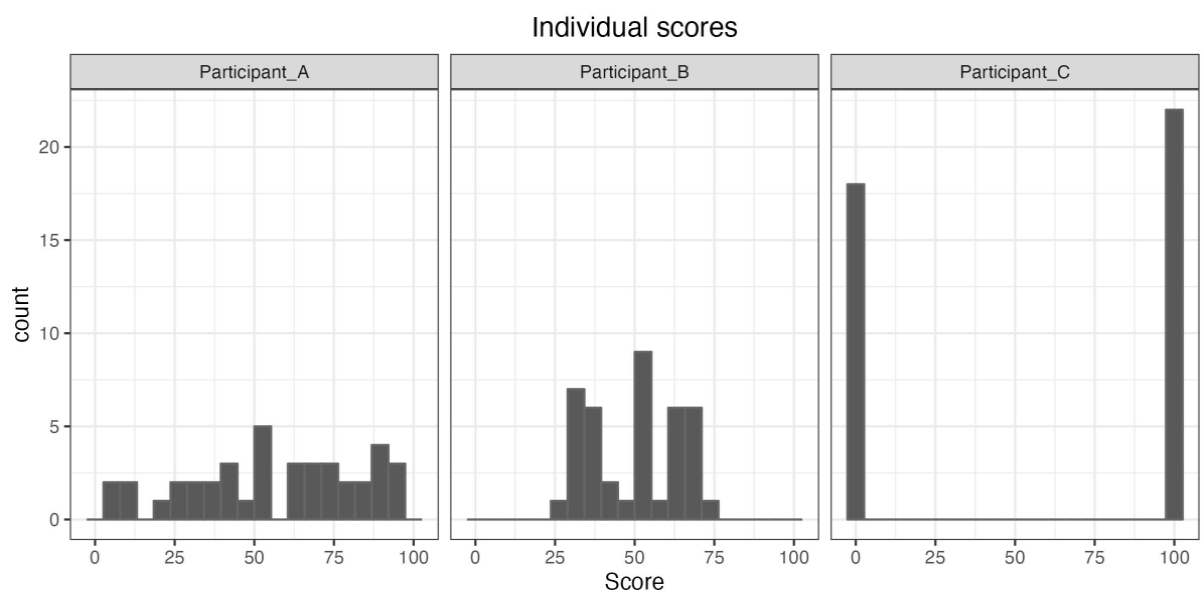

Figure 9. Individual variability in using a 100-point slider scale in an experiment.

The *brms* package also supports modeling zero-inflated data – cases where there are many zeros, but few ones. For example, Wirtz et al. (2024) analysed data from the Atlas of Colloquial German, where the response variable was the diversity of linguistic variants known to each informant. In many cases, informants knew only one variant, which corresponded to the diversity score of 0. For such data, zero-inflated regression was an appropriate approach. To summarize, ZOIB regression and its variants offer a flexible framework for modelling slider data and other datasets with a high concentration of observations at one or both extremes of the scale.

4.4. The problems of too many predictors ("large P") and multicollinearity

In some linguistic models, particularly those based on corpus data, it is common to have a large number of predictors. This situation often arises in supervised machine learning applications, for instance when using dimensions from word embeddings to predict a linguistic variable. Including all predictors without selection can result in overly complex models.

In the frequentist framework, common solutions include ridge regression, LASSO regression, and the elastic net, which represents a combination of both (for linguistic applications, see, e.g., Tomaschek et al. 2018; Van de Velde & Pijpops 2019). These approaches can be interpreted in Bayesian terms as models with shrinkage priors: LASSO corresponds to a Bayesian model with Laplacian priors, which have a sharp peak at zero, while ridge regression corresponds to a model with normal priors. Another type of shrinkage priors is the Horseshoe prior, introduced by Carvalho et al. (2009) for datasets with many predictors whose coefficients are zero or near zero.[2]

Shrinkage priors can also be useful in cases of (multi)collinearity, where predictors are strongly correlated. Multicollinearity does not necessarily reduce a model's overall predictive performance, but it often makes the coefficients of the individual predictors difficult or even impossible to identify. This can lead to unreliable estimates, increased standard errors, increased Type I error rates and other issues (e.g., Kalnins 2018; Tomaschek et al. 2018). Shrinkage priors can alleviate these problems (Rashid et al. 2022; Tzoumerkas & Fouskakis 2024).

Another valuable approach is Bayesian Model Averaging (BMA). When working with many predictors, researchers often perform model selection and then present the final model as if the model had been specified in advance. This ignores the uncertainty inherent in the process of model selection, and the possibility that several competing models could plausibly have generated the data (Kaplan 2021). BMA addresses this by averaging over models rather than selecting a single "best" one. One method is Markov Chain Monte Carlo Model Composition (MC³). Instead of exploring only the space of possible parameter values, as in standard Bayesian inference, MC³ performs a random walk over the space of possible models. This yields posterior model probabilities (PMP), which can be used to identify the highest-scoring models.

If the aim is prediction, PMPs can be used for weighting the outcome predictions made by different models. If the goal is to identify relevant predictors, BMA can produce posterior inclusion probabilities, the proportion of visited models in which each predictor appears. One can also compute the proportion of visited models in which a = coefficient is positive or negative (see illustrations in Kaplan 2021). Various extensions of model averaging, such as model stacking, are available in *brms*, *BMA* (Raftery et al. 2025) and *BMS* (Zeugner & Feldkircher 2015).

---

[2] The name 'horseshoe prior' comes from a symmetric horseshoe-shaped profile of the shrinkage factor under specific values of parameters. The shrinkage factor has a very high probability density in two extremes: no shrinkage for very strong signals and total shrinkage for zero signals (Carvalho et al. 2009).

4.5. Small sample size ("small N" problem)

In some areas of linguistics, data collection is costly – for example, in studies based on spoken and multimodal corpora, investigations of rare typological phenomena, and certain experimental designs. A common challenge when fitting Maximum Likelihood (ML) models on small samples is that parameter estimates can become unreasonably extreme. Another risk is overfitting. Incorproating prior information can help mitigate these problems.

Levshina (2022) illustrates this with a study of variation between *help* + bare infinitive and *help* + *to*-infinitive in English using a set of contextual predictors. She first fitted both ML and Bayesian models with weak regularizing priors to a large dataset, obtaining convergent results. To test an additional theoretical hypothesis, she introduced one more predictor, but the manual coding required was time-consuming, so only a small dataset could be annotated. When an ML model was fitted to this smaller dataset, it overfit severely, producing extreme estimates with wide confidence intervals, rendering the model essentially useless. The problem was addressed by fitting a Bayesian model with strong informative priors: normal distributions whose means and standard deviations matched the posterior distributions from the large-sample model. For the new predictor, weak regularizing priors were used. This Bayesian model replicated the effects of the ML and Bayesian large-sample models for the shared predictors, while showing no convincing effect of the new variable on the linguistic variation.

4.6. Bayesian meta-analysis

Another Bayesian method potentially useful to linguists is Bayesian meta-analysis. This method accumulates evidence across numerous studies, estimating a pooled effect in a Bayesian hierarchical model. For example, Vasishth et al. (2018) used Bayesian meta-analysis to assess whether incomplete neutralization effects in German final devoicing are robust. This approach is valuable when individual studies have small sample sizes, which makes reliable statistical inference difficult. Bayesian meta-analysis can also be superior when the number of included studies is small because it directly models the uncertainty in between-study heterogeneity.

4.7. Testing the null hypothesis

Finally, Bayesian methods can be used, as Dienes puts it candidly (2014), to "get the most out of nonsignificant results". In a Maximum Likelihood regression, a *p*-value greater than 0.05 does not reveal whether this is because the null hypothesis is true, or because the dataset lacks sufficient power. By contrast, Bayes Factor can quantify the evidence for the null hypothesis relative to the alternative, offering a more informative assessment.

For example, De Rosa et al. (2024) used Bayes Factor to test if brain responses differed between meaningful suffixes vs. meaningless word endings in a reading task. Across multiple regions of interest, they found mostly moderate evidence supporting the null hypothesis of no difference. This finding has important theoretical implications, lending support to cognitive models of morphological processing that emphasize perceptual and orthographic mechanisms in the decomposition of complex words.

The Bayesian approach can help address the issue of so-called null results – research findings that do not support a tested hypothesis, – which remain severely underreported despite their importance for scientific progress. Such results can inspire new lines of inquiry, reduce research waste and improve transparency.[3] At the same time, Bayes Factor is not magic. First of all, if there is too little data and/or the effect sizes are small, Bayes Factor will not solve these problems. The results are likely to be inconclusive (Schad et al. 2023).

5. **Perspectives, challenges and good practices**

Although Bayesian statistics is still far from mainstream in linguistics, its methods are increasingly being adopted, particularly in areas that demand complex multivariate hypothesis testing. Applications can be found across all methodologies and across virtually all linguistic subfields.

But do we really need to choose between frequentist and Bayesian methods? Dienes (2011) asks a provocative question in his paper, "Bayesian Versus Orthodox Statistics: Which

---

[3] See Springer Nature's white paper: https://stories.springernature.com/the-state-of-null-results-white-paper/executive-summary/index.html (last access 18.08.2025).

Side Are You On?", but this is not the most productive way of approaching the problem, in my view. Instead, I concur with Roettger et al. (2019: 3):

"A spirit of plurality is needed that creates space for realizing that there are problems and applications that might be handled more easily by either Bayesian or frequentist approaches, and that the analyst is far better off having both tools in their toolbox. In particular, researchers need to be familiar with both approaches, since there is an increasing number of papers in quantitative linguistics that uses Bayesian approaches."

Both approaches have their strengths and weaknesses, but the appeal of Bayesian statistics is especially strong in contexts where traditional approaches struggle: handling non-Gaussian response variables, conducting meta-analyses, working with small datasets, or modeling complex random effects structures. Moreover, Bayesian methods are uniquely suited for quantifying evidence in support of the null hypothesis.

At the same time, Bayesian methods in linguistics present quite a few challenges. Because their use is still relatively new, there is no consensus of best practices. The absence of strict conventions can also lead to uncertainty and a sense of "choice overload", especially for beginners. For example, while most researchers report mean posterior values, others prefer medians. Criteria for determining the presence or absence of an effect vary, as well: some use the Probability of Direction (PD), others the Region of Practical Equivalence (ROPE), and still others rely on credible intervals (which are, however, not the most suitable tool for making such statements). Moreover, there is no agreements about which credible intervals to report. Some researchers use Equal-tailed intervals (ETI), while others prefer Highest Density Interval (HDI). The cut-offs used by different authors vary, as well: 50%, 66%, 80%, 89%, 90%, 95% or 99%. Approaches to model diagnostics are equally diverse. For example, some insist on R-hat values of exactly 1.00, while others accept values below 1.05, and so on.

While Bayesian statistics in many cases is more transparent than frequentist statistics (for example, in the interpretation of credible intervals), some Bayesian concepts can also be misleading if misapplied. A notable example is Bayes Factor, which is frequently used and interpreted incorrectly (Tendeiro & Kiers 2019). An important limitation of Bayes Factors is their sensitivity to priors, which enter into the computation of the model's likelihood. If the priors are too diffuse or poorly chosen, Bayes Factors can yield unrealistic values. This is why some statisticians remain sceptical of their usefulness (Gelman et al. 2003).

Moreover, just as the misuse of *p*-values has given rise to p-hacking, Bayesian statistics is not immune to similar problems. A practice sometimes called "BF-hacking" (Konijn et al. 2015) exploits the conventional cut-off values to interpret Bayes Factors. Following Jeffreys (1961), a BF between 1/3 and 3 is regarded as "anecdotal evidence", a BF between 3 and 10 or between 1/3 and 1/10 as "moderate", and values above 10 or below 1/10 as "strong". Yet, a BF of 3.01 should not be considered substantially different from a BF of 2.99. Konijn et al. (2015) use the famous quote of Rosnow and Rosenthal: ". . . surely, God loves the .06 nearly as much as the .05" (Rosnow and Rosenthal 1989: 1277), in the Bayesian context: "God would love a Bayes Factor of 3.01 nearly as much as a BF of 2.99."

All this means that it may be the right moment to reflect on what constitutes good practices in Bayesian statistics. I would like to conclude this paper with several proposals.

1. **Think carefully about your priors**. Priors can be informed by elicitation of information from experts, previous studies, or common sense. Although some general priors have been proposed, such as Cauchy(0, 2.5) for standardized input variables in logistic regression, such defaults should be regarded as useful placeholders, which are outperformed by more specific priors (Gelman et al. 2008). Before fitting your regression model, conduct prior predictive checks to ensure that your priors are sensible and correctly specified. The basic idea is to simulate vectors of regression coefficients from the prior, generate predicted outcomes based on those coefficients, and verify that the outcomes are similar to the actual data. When fitting your model, also perform a prior sensitivity analysis, trying out different possible priors to see if they affect the posterior inferences. This step is especially important if your dataset is small. Describe all the priors you used, even if they are software defaults, since default settings may change over time.
2. **Check chain convergence**. Make sure that your Markov chains mix well and converge to a stationary distribution. Use standard diagnostics, such as R-hats and effective sample sizes, and visual inspection to assess convergence. If necessary, increase the number of iterations or adjust other control parameters. When done, report key details of the sampling procedure, such as the number of chains, the length of the warm-up period and the total number of iterations per chain.

3. **Be efficient**. MCMC sampling takes substantial time and computational power. In the early stages of modelling, it is sensible to run fewer and shorter chains, provided that the R-hat values and standard errors are acceptable. Longer MCMC chains should be used at later stages (Gelman et al. 2020). Make use of parallelism, running separate chains on multiple cores. When a rough approximation to the posterior distribution is sufficient, one can use alternative algorithms, such as variational inference or Laplace approximation.

4. **Report the posterior distributions comprehensively.** Provide detailed information about the posterior distributions. Always report credible intervals, ideally at several different quantile levels, e.g., 69%, 89% and 97% (McElreath 2016). Inspect their full distributional shape to avoid multimodal distibutions. If the posterior distribution is asymmetric, use the median and HDI rather than ETI. Focus on the magnitude and uncertainty of effects, avoiding the binary "effect/no effect" decisions.

5. **Perform model diagnostics**. Bayesian regression is not a magic bullet. Bayesian models can still suffer from overfitting, heteroscedasticity and misspecification, and these issues should be addressed with the same care as in frequentist regression analysis. One important procedure is performing posterior predictive checks. Generate simulated data conditional on the posterior distribution and compare these simulations to the observed data. This helps assess whether the model adequately captures the patterns in the data.

6. **Use Bayes Factors with caution**. If you rely on Bayes Factors, select appropriate informative priors (Schad et al. 2023), ideally derived from other empirical studies with a similar design. Perform sensitivity analyses to evaluate how the choice of different priors affects Bayes Factor values. Ensure that your MCMC chains are long enough to provide sufficient effective sample sizes.